\documentstyle[12pt,amssymb,amsmath]{article}
\begin{document}

{\Large

\noindent {\bf Polynomially deformed oscillators as $k$-bonacci
\\ oscillators}}

\bigskip

\noindent{\bf A M Gavrilik and A P Rebesh}

\noindent Bogolyubov Institute for Theoretical Physics, Kiev 03680,
Ukraine

\noindent E-mail: omgavr@bitp.kiev.ua

\medskip

\begin{abstract}
A family of multi-parameter, polynomially deformed oscillators
(PDOs) given by polynomial structure function $\varphi(n)$ is
studied from the viewpoint of being (or not) in the class of
Fibonacci oscillators. These obey the Fibonacci relation/property
(FR/FP) meaning that the $n$-th level energy $E_n$ is given
linearly, with real coefficients, by the two preceding ones
$E_{n-1}$, $E_{n-2}$. We first prove that the PDOs do not fall in
the Fibonacci class. Then, three different paths of generalizing the
usual FP are developed for these oscillators: we prove that the PDOs
satisfy respective $k$-term generalized Fibonacci (or "$k$-bonacci")
relations; for these same oscillators we examine two other
generalizations of the FR, the inhomogeneous FR and the
"quasi-Fibonacci" relation. Extended families of deformed
oscillators are studied too: the $(q;\mu)$-oscillator with
$\varphi(n)$ quadratic in the basic $q$-number $[n]_q$ is shown to
be Tribonacci one, while the $(p,q;\mu)$-oscillators with
$\varphi(n)$ quadratic (cubic) in the $p,\!q$-number $[n]_{p,q}$ are
proven to obey the Pentanacci (Nine-bonacci) relations. Oscillators
with general $\varphi(n)$, polynomial in $[n]_{q}$ or $[n]_{p,q}$,
are also studied.
\end{abstract}





 \bigskip



\bigskip

\section{Introduction}\hspace{5mm}
As it is known for some time, among various non-linear
generalizations or deformations of the usual quantum harmonic
oscillator there is a distinguished class of so-called Fibonacci
oscillators \cite{Arik1} - the oscillators whose energy spectra
satisfy the Fibonacci property (FP), implying: $E_{n+1}=\lambda
E_n+\rho E_{n-1}$, with real constants\!\! \footnote {The famous
Fibonacci numbers stem from the relation $F_{n+1}=F_{n}+ F_{n-1}$
where $F_0=F_1=1$.} $\lambda$ and $\rho$.
 As stated in \cite{Arik1}, the Fibonacci class is just 
the two-parameter deformed family of
$p,\!q$-oscillators, introduced in \cite{CJ}.
 The family of $p,\!q$-oscillators is rich enough.
 In particular, it contains such exotic one-parameter $q$-oscillator
as Tamm-Dancoff (TD) deformed oscillator \cite{Odaka}, \cite{Jagan}
which possesses besides the FP a set of nontrivial properties as
shown in \cite{GR1}.
 Moreover, a whole plenty of different one-parameter deformed
oscillators are contained in this family as particular cases.
 Most of them, except for the best known $q$-oscillators of
Arik-Cook (AC) \cite{Arik2} and Biedenharn-Macfarlane (BM)
\cite{BM}, are not well-studied but nevertheless have some potential
\cite{GR2} for possible applications. What concerns the
$p,\!q$-deformed Fibonacci oscillators, there exist some rather
unusual properties and already elaborated interesting physical
applications, see \cite{Ch-J}-\!\cite{GR-UJP}.
 However, a natural question arises whether the family of
$p,\!q$-oscillators exhausts the Fibonacci class. In that
connection, recently we have shown in Ref. \cite{GKR} that definite
3-, 4-, and 5-parameter deformed extensions of the
$p,\!q$-oscillator considered in \cite{Chung-PhL, Borzov, Mizrahi,
Burban} also belong to the Fibonacci class, i.e., possess the FP.
 In that same paper, we studied a principally different,
so-called $\mu$-deformed
oscillator proposed earlier in \cite{Jann}, and shown
that it {\it does not possess} the FP.
 For that reason, a new concept has been developed for this
$\mu$-oscillator. Namely, it was demonstrated that the
$\mu$-oscillator belongs to the more general, than Fibonacci,
class of so-called "quasi-Fibonacci"\ oscillators \cite{GKR}.

The goal of the present paper is to study yet another classes of
nonlinear deformed oscillators which do not belong to the Fibonacci
class.
We treat, from the viewpoint of three possible ways of
generalizing the FP, a class of polynomially deformed oscillators.
 It is proven, using the notion of deformed oscillator structure function
\cite{Mel,Man',Bona}, that those oscillators are principally of
non-Fibonacci nature.
 Then we develop the generalization of FP for these oscillators
along three completely different paths: (i) as oscillators with
$k$-term generalized Fibonacci property; (ii) as oscillators obeying
inhomogeneous Fibonacci relation; (iii) as quasi-Fibonacci
oscillators.
 Besides, we study a family of $(q;{\mu})$-oscillators which is,
in a sense, a mix of the quadratic and the AC type $q$-deformed
oscillators, and demonstrate its Tribonacci property.
 This result is extended to a general $r$-th order polynomial in the
AC-type of $q$-oscillator bracket $[N]_q$, naturally leading to
$k$-bonacci relations. In this respect, let us mention that similar
$k$-bonacci relations were treated in \cite{Schork} in connection
with generalized Heisenberg algebras \cite{Souza}. Likewise, for the
$(q;\{\mu\})$-oscillators with $\{\mu\}\!=\!(\mu_1,\mu_2,...,
\mu_r)$, combining the polynomial and the $q$-deformed AC features,
the general statement on their $k$-bonacci property is proven. In a
similar manner, the three parameter $(p,q;\mu)$-deformed oscillators
are treated as well and shown to obey their characteristic
Pentanacci property. Finally, for certain four-parameter or
$(p,q;\mu_1,\mu_2)$-deformed family of nonlinear oscillators we
demonstrate the validity of Nine-bonacci relation by finding
explicitly the relevant nine coefficients $A_j(p,q)$.

\section{Polynomially deformed or $\{\mu\}$- oscillators}\hspace{5mm}
In the preceding work \cite{GKR} we have shown that the
$\mu$-oscillator from \cite{Jann} does not satisfy the usual (with
{\it two-term} RHS) linear, homogeneous Fibonacci relation (FR)
\begin{equation}
E_{n+1}=\lambda E_n+\rho E_{n-1}\ , \label{1}
\end{equation}
with $\lambda$ and $\rho$ some real, constant coefficients. To make
the $\mu$-deformed oscillator from \cite{Jann} satisfy a relation
like (\ref{1}), the important modification is needed: the
coefficients should depend on $n$: $\lambda\!=\!\lambda(n)$,
$\rho\!=\!\rho(n)$ (i.e., not constants).
 That is, this way of modifying the FP involves the coefficients,
not the shape of relation.
 What concerns polynomially deformed oscillators to be studied here,
we will demonstrate that they admit three different approaches for
generalizing the FP.

Like in \cite{Mel,Man'}, we study the algebra of deformed oscillator
through its {\em structure function}: $a^{\dagger}a\!=\!\varphi(N)$
and $aa^{\dagger}\!=\!\varphi(N+1)$.
 Note, the same structure function
determines both the basic commutation relation of $a, a^{\dagger}$
and the Hamiltonian and energy eigenvalues:
\begin{equation}
aa^{\dagger}-a^{\dagger}a=\varphi(N+1)-\varphi(N), \label{2}
\end{equation}
\begin{equation}
H=\frac12\Bigl(\varphi(N+1)+\varphi(N)\Bigr), \hspace{18mm}
E_n=\frac12\Bigl(\varphi(n)+\varphi(n+1)\Bigr)\ .\label{3}
\end{equation}
The latter formula implies usage of the properly modified version of
Fock space wherein (see e.g., \cite{Bona})
\begin{equation}
a|0\rangle = 0, \hspace{5mm}  N |n\rangle = n|n\rangle, \hspace{5mm}
\varphi(N)|n\rangle = \varphi(n)|n\rangle, \hspace{5mm}  H|n\rangle
= E_n|n\rangle .
\label{4}
\end{equation}

In this paper we focus on the polynomially deformed oscillator.
 Its structure function
\begin{equation}
\varphi(N)=N+\sum_{i=1}^{r}\mu_iN^{i+1}, \hspace{18mm} \mu_i\geq 0\
, \label{5}
\end{equation}
involves the parameters $\{\mu\}\equiv(\mu_1, \mu_2,...,\mu_r)$, so
these polynomial oscillators may also be termed the
$\{\mu\}$-deformed ones. Note, the restriction on $\mu_i$ provides
positivity and monotonicity of the energies $E_n$ in (\ref{3}). It
is worth to remark that for the $\{\mu\}$-deformed oscillator given
by (\ref{5}), the basic relation can be presented, instead of
(\ref{2}), also as
\begin{equation}
aa^{\dagger}-qa^{\dagger}a=f(N)=\sum_{l=0}^{r+1} \alpha_lN^l,
\hspace{18mm} \alpha_l\in\mathbf{R}.
 \label{6}
\end{equation}
We can translate the form (\ref{2}) of basic relation into the
latter one (\ref{6}). Indeed, taking the $q$-commutator of $a$ and
$a^{\dagger}$ for the deformed $\{\mu\}$-oscillator we have (we set
$\mu_0=1$)
\[
aa^{\dagger}-qa^{\dagger}a=\varphi(N+1)-q\varphi(N)=
\]
\[
=N+1+\sum_{j=1}^{r}\mu_j(N+1)^{j+1}-
q\Bigl(N+\sum_{j=1}^{r}\mu_jN^{j+1}\Bigr)=
\]
\begin{equation}
=\sum_{j=0}^{r}\mu_j\biggl(- q
N^{j+1}+\sum_{s=0}^{j+1}\frac{(j+1)!}{s!(j+1-s)!}N^s\biggr).
\label{7}
\end{equation}
The latter relation goes over into (\ref{6}) if
\begin{equation}
\alpha_0=\mu_0+\mu_1+\mu_2+...+\mu_{r}=\sum_{s=1}^{r+1}\mu_{s-1}
\label{8}
\end{equation}
\begin{equation}
\alpha_l=-q\mu_{l-1}+\sum_{s=1}^{r+1}\frac{s!}{l!(s-l)!}\mu_{s-1},
\hspace{12mm}  1\leq l\leq r+1 . \label{9}
\end{equation}
The form of basic relation similar to (\ref{6}) was used in
\cite{Chung-JMP} to treat the polynomial oscillators.


\subsection{Non-Fibonacci nature of polynomial $\{\mu\}$-oscillators}\hspace{5mm}
Let us first demonstrate that the polynomially deformed oscillators,
see (\ref{5}), do not satisfy the relation (\ref{1}) if one insists
on the constant nature of its coefficients.

Usual quantum harmonic oscillator which has $\varphi(n)=n$ and the
linear energy spectrum $E_n=\frac12(2n+1)$, is just the particular
$r=0$ case of (\ref{5}). As is known, this oscillator with
$\lambda=2$ and $\rho=-1$ satisfies the standard FR (\ref{1}) .
 Such property, however, fails if $r=1$, i.e., for the quadratic,
with $\varphi(n)=n+\mu_1n^2$, deformation of harmonic oscillator
cannot satisfy the standard FR (\ref{1}).

The FR (1) fails also for the cubic $r=2$ extension with
$\varphi(n)=n+\mu_1 n^2+\mu_2 n^3$ for which the energy spectrum is
$E_n=\frac{1}{2}(n+\mu_1 n^2+\mu_2
n^3+n+1+\mu_1(n+1)^2+\mu_2(n+1)^3)$. To show the failure, we insert
the cubic $\varphi(n)$ into (\ref{1}) and deduce the system of
equations ($\mu_1,\mu_2\ne 0$):
\[n^3: \ \ \ \ \mu_2-\rho\mu_2-\lambda\mu_2 = 0\ ;\]
\begin{equation}
n^2: \ \ \ \
\hspace{-3mm}-\frac{3}{2}\lambda\mu_2-\rho\mu_1+\frac{9}{2}\mu_2+\mu_1+
          \frac{3}{2}\rho\mu_2-\lambda\mu_1
= 0\ ; \label{10}
\end{equation}
\[n^1: \ \ \ \
1+\rho\mu_1+3\mu_1-\frac{3}{2}\lambda\mu_2-\lambda-\lambda\mu_1+
\frac{15}{2}\mu_2-\frac{3}{2}\rho\mu_2-\rho = 0\ ;\]
\[n^0: \ \ \ \
\frac{1}{2}\rho-\frac{1}{2}\rho\mu_1+\frac{5}{2}\rho+\frac{9}{2}\mu_2-
 \frac{1}{2}\lambda-\frac{1}{2}\lambda\mu_1-
\frac{1}{2}\lambda\mu_2+\frac{1}{2}\rho\mu_2+\frac{3}{2}=0\ .\]

\noindent The top two equations are solved with $\lambda=2$ and
$\rho=-1$, but, these values are incompatible with the rest of
equations in the system, that proves the statement.

One can prove for general situation that the $r$-th order
polynomially deformed oscillator (the structure function is of the
order $r\geq2$) does not satisfy the standard FR (\ref{1}).
 Again, the equations got at two senior powers of $n$ yield
$\lambda=2$ and $\rho=-1$ as solution, but these values are
incompatible with the rest of equations in the system.

Since the FP fails for polynomial $\varphi(n)$, we
consider possible extensions of the FP.

\vspace{3mm}

\subsection{A $k$-term extended ($k$-bonacci) oscillators}\hspace{5mm}
We begin with quadratic oscillator and extend the FR by adding one
term:
\begin{equation}
E_{n+1}=\lambda_0 E_n+\lambda_1 E_{n-1}+\lambda_2 E_{n-2}\ .
\label{11}
\end{equation}
 This is the {\it three-step generalized} Fibonacci
or Tribonacci (see  e.g. \cite{Schork}) relation.

As $\lambda_0$, $\lambda_1$, $\lambda_2$ are constants, and in view
of (\ref{3}), it is sufficient to deal with the relation
\begin{equation}
\varphi_{n+1}=\lambda_0 \varphi_n+\lambda_1 \varphi_{n-1}+\lambda_2
\varphi_{n-2}\ , \hspace{18mm} \varphi_n\equiv\varphi(n). \label{12}
\end{equation}
Indeed, if (\ref{12}) is valid the relation (\ref{11}) is valid too.
So, insert in (\ref{12}) the quadratic $\varphi(n)$ that is the
$r=1$ case of (\ref{5}). Solving the system of equations deduced
similarly to (\ref{10}) we find $\lambda_0=3$, $\lambda_1=-3$,
$\lambda_2=1$. With these coefficients we verify that the relation
(\ref{11}) does hold.

Now consider  general case of polynomially deformed oscillators
given by  the structure function (\ref{5}),
with any $r\geq1$.
 Accordingly, consider the $k$-term extension of FR,
or $k$-bonacci relation,
of the form ($n\geq k-1$) \vspace{-2mm}
\begin{equation}
E_{n+1}=\lambda_0 E_n+\lambda_1 E_{n-1}+\lambda_2
E_{n-2}+...+\lambda_{k-1}E_{n-k+1}=\sum_{i=0}^{k-1}\lambda_i
E_{n-i}\ . \label{13}
\end{equation}
 Then the following statement is true.

{\bf Proposition 1.} The energy values $E_n$, given by (\ref{3}) of
the polynomially deformed oscillator with structure function
(\ref{5}) satisfy the $k$-generalized FR (\ref{13}) if $r=k-2$ and
$\lambda_i$ are given as\footnote{Since the set of coefficients
$\lambda_i$ of (\ref{14}) obviously depend on fixed $k$, we will
indicate this explicitly.}
\begin{equation}
\lambda^{(k)}_i=(-1)^i\frac{k!}{(i+1)!(k-1-i)!}=(-1)^i\left(
       \begin{array}{c}
              \hspace{-2mm}k\hspace{-2mm}\\
                                 \hspace{-2mm}i\!+\!1\hspace{-2mm}\\
                                  \end{array}
                                    \right). \label{14}
\end{equation}

{\it Proof.} Clearly, the $k$-term relation (\ref{13}) will be valid
for the energy values if the structure function given in (\ref{5})
with $r=k-2$ satisfies the same equality, written as
\begin{equation}
n+1+\sum_{j=1}^{k-2}\mu_j(n+1)^{j+1}-\sum_{i=0}^{k-1}
  \lambda^{(k)}_i
\biggl(n-i+\sum_{j=1}^{k-2}\mu_j(n-i)^{j+1}\biggr)=0\ . \label{15}
\end{equation}
The latter will be proven by induction.
 Supposing that the $(k-1)$-term relation
\begin{equation}
n+1+\sum_{j=1}^{k-3}\mu_j(n+1)^{j+1}-\sum_{i=0}^{k-2}\lambda_i^{(k-1)}
\biggl(n-i+\sum_{j=1}^{k-3}\mu_j(n-i)^{j+1}\biggr)=0 \label{16}
\end{equation}
holds for the structure function $
\varphi(n)\!=\!n\!+\!\sum_{i=1}^{k-3}\mu_in^{i+1} $ with
$\lambda_i^{(k-1)}$ as in (\ref{14}),
we then prove that the structure function
 $\varphi(n)=n+\sum_{i=1}^{k-2}\mu_in^{i+1}$ satisfies the
relation (\ref{15}) with $\lambda_i^{(k)}$ from (\ref{14}).
 But, first let us  check that (\ref{15}) along with (\ref{14}) is
true for $k=2,3$.

If $k=2$ that means the usual linear quantum oscillator, the
relation is just the standard 2-term FR
$E_{n+1}=\lambda_0E_n+\lambda_1E_{n-1}$: it does hold for
$\lambda_0=2$ and $\lambda_1=-1$ since for these $\lambda_0,\
\lambda_1$ the following pair of relations
\[  \varphi_{n+1}=\lambda_0\varphi_n+\lambda_1\varphi_{n-1},
\hspace{12mm}
\varphi_{n}=\lambda_0\varphi_{n-1}+\lambda_1\varphi_{n-2} ,
\]
is obviously true, which read:
\begin{equation}
n+1-2n-(-1)(n-1)=0, \hspace{12mm} n-2(n-1)-(-1)(n-2)=0. \label{17}
\end{equation}
If $k=3$ or for quadratic $\varphi_n=n+\mu_1n^2$,
we have the Tribonacci relation
\begin{equation}
\varphi_{n+1}=\lambda_0\varphi_n+\lambda_1\varphi_{n-1}+
\lambda_2\varphi_{n-2}. \label{18}
\end{equation}
It rewrites in the form (\ref{15}), that is
\[
n+1+\mu_1(n+1)^2=3(n+\mu_1
n^2)+(-3)(n-1+\mu_1(n-1)^2)+1(n-2+\mu_1(n-2)^2)
\]
where $\lambda_0=3$, $\lambda_1=-3$ and $\lambda_2=1$. The latter
relation, with account of the both identities in (\ref{17}), reduces to
\[
\mu_1(n+1)^2-3\mu_1n^2+3\mu_1(n-1)^2-\mu_1(n-2)^2=0
\]
where we encounter the full squares only. This, as easily checked,
holds identically.

Similar reasonings are applied to the situation of general
polynomial $\varphi(n)$. Note first that $\lambda_i^{(k)}$ in
(\ref{14}) split as
\begin{equation}
\lambda_i^{(k)}=\lambda_i^{(k-1)}-\lambda_{i-1}^{(k-1)}. \label{19}
\end{equation}
Using this splitting in the LHS of the $k$-th order generalized
Fibonacci ($k$-bonacci) relation (\ref{15}) we extract {\it twice}
the (supposed to hold) $(k-1)$-term generalized FR (\ref{16}): first,
in the form of LHS of (\ref{16}), for fixed $n$, with
$\lambda_i^{(k-1)}$ involved and, second, in the form of LHS of
(\ref{16}) rewritten for $n\rightarrow n-1$ and involving the set
$(-1)\lambda_{i-1}^{(k-1)}$ from (\ref{19}). As result, we get the
relation consisting of the highest $(k-1)$-th order terms (in $n+1$ or in
$n-i$) only:
\begin{equation}
\mu_{k-1}(n+1)^{k-1}-\sum_{i=0}^{k-1}\lambda_i^{(k)}
\bigl(\mu_{k-1}(n-i)^{k-1}\bigr)=0. \label{20}
\end{equation}
Since $\mu_{k-1}\neq 0$, the latter relation rewrites as
\begin{equation}
F_n(k,\lambda_i^{(k)})\equiv(n+1)^{k-1}-\sum_{i=0}^{k-1}\lambda_i^{(k)}
\bigl(n-i\bigr)^{k-1}=0. \label{21}
\end{equation}
Then, to prove (\ref{21}), we expand the binomials and interchange
the summation order:
\[
F_n(k,\lambda_i^{(k)})=\sum_{s=0}^{k-1}\frac{(k-1)!}{s!(k-1-s)!}~n^{k-1-s}1^s-
\]
\[
-\sum_{i=0}^{k-1}(-1)^i\frac{k!}{(i+1)!(k-1-i)!}
\sum_{s=0}^{k-1}\frac{(k-1)!}{s!(k-1-s)!}~n^{k-1-s}(-1)^si^s=
\]
\[
=\sum_{s=0}^{k-1}\frac{(k-1)!}{s!(k-1-s)!}~n^{k-1-s}\biggl(1-(-1)^sk!
\sum_{i=0}^{k-1}(-1)^i\frac{i^s}{(i+1)!(k-1-i)!}\biggr)=
\]
\[
=\sum_{s=0}^{k-1}\frac{(k-1)!}{s!(k-1-s)!}n^{k-1-s}\biggl((-1)^{s+1}
\sum_{-1\leq i\leq k-1}(-1)^{i}\frac{k!}{(i+1)!(k-1-i)!}~i^s\biggr)
\]
(note that the entity $1$ is included in the sum as the additional
$i=-1$ term).

Shifting the index $i$ as $i\to i-1$ we obtain
\[
F_n(k,\lambda_i^{(k)})=\sum_{s=0}^{k-1}\frac{(k\!-\!1)!}{s!(k\!-\!1\!-\!s)!}
n^{k-1-s}\biggl((-1)^{s+1}\!\sum_{-1\leq i-1\leq
k-1}(-1)^{i-1}\frac{k!}{i!(k-i)!}(i\!-\!1)^s\biggr)\!=
\]
\[
=\sum_{i=0}^{k-1}\frac{(k-1)!}{s!(k-1-s)!}n^{k-1-s}\biggl((-1)^s
\sum_{i=0}^{k}(-1)^{i}\frac{k!}{i!(k-i)!}(i-1)^s\biggr)=0\
\]
where the fact of final turning into zero is due to the formula
\[
\sum_{j=0}^{k}(-1)^j\frac{k!}{j!(k-j)!}(j-1)^m=0, \hspace{8mm}
m=0,1,2,...,k-1\ ,
\]
which can be proven analogously to the known formula        \cite{Korn}
\[
\sum_{j=0}^k(-1)^j\frac{k!}{j!(k-j)!}~j^m=0, \hspace{8mm}
m=0,1,2,...,k-1\ .
\]
 Thus we gain the proof.

{\bf Remark 1.} It is remarkable that the set (\ref{14}) of the
coefficients $\lambda_i^{(k)}$\hspace{-1mm}, $i\!=\!0,1,2,...,k\!-\!1$, which provide
the validity of the $k$-term Fibonacci relation (\ref{13}) for the
polynomially deformed oscillators with the structure function
 $\varphi(n)=n+\sum_{i=1}^{k-2}\mu_in^{i+1}$, see (\ref{5}),
are {\em totally independent} of the parameters $\mu_i$ of
$\varphi(n)$.
 In particular, some of the $\mu_i$ (but not the "senior"\ one $\mu_{k-2}$)
may be equal to zero.

 {\bf Remark 2.} The content of the
Proposition 1 can be extended to the cases $r\!<\!k\!-\!2$ or
$r\!>\!k\!-\!2$. Namely, it can be demonstrated that the $k$-term
Fibonacci relation is satisfied for all the polynomial oscillators
for which $r\!<\!k\!-\!2$. Equivalently, the oscillator with $r$-th
order polynomial structure function satisfies all the $k$-term
generalized FR such that $k>r+2$, with appropriate coefficients.
 For instance, the quadratic oscillator which obeys the 3-term or Tribonacci
relation, see (\ref{18}), with fixed $\lambda_0$, $\lambda_1$ and
$\lambda_2$ equal respectively to 3, -3, 1, obviously satisfies,
with definite four coefficients, also the 4-term relation
\[
\varphi_{n+1}=(\lambda_0-1)\varphi_n+(\lambda_0+\lambda_1)\varphi_{n-1}+
(\lambda_1+\lambda_2)\varphi_{n-2}+\lambda_2\varphi_{n-3}\ ,
\]
and with proper coefficients also the higher order 5-term, 6-term,
etc., $k$-bonacci relations.
 On the other hand, the oscillator with $r$-th
order polynomial structure function does not satisfy any $k$-term
generalized Fibonacci relations such that $k<r+2$. Accordingly, the
$k$-term Fibonacci relation is not valid for those polynomial
oscillators for which $r>k-2$.

\vspace{3mm}

\vspace{0.2cm}
  \subsection{\hspace{-2mm} Polynomial $\{\mu\}$-oscillators: inhomogeneous
FR}\hspace{5mm}
 Here we consider an alternative (though also linear in the energy eigenvalues)
form of generalized
FR which is valid for the polynomially deformed oscillators:
\begin{equation}
E_{n+1}=\lambda E_n +\rho E_{n-1} +\sum_{i=0}^{k-1}\alpha_{i}n^i,
\label{22}
\hspace{15mm} E_n=\frac12\bigl(\varphi(n)+\varphi(n+1)\bigr).
\end{equation}
For obvious reason and in analogy with \cite{Asveld}, we call such an extension of
FR the "inhomogeneous Fibonacci relation".
 Again it is sufficient to deal, instead of the energy by itself, with the structure
function (\ref{5})
of deformed oscillator.
 So, consider two relations
\begin{equation}
\varphi(n\!+\!1)\!=\!\lambda \varphi(n)+\rho\, \varphi(n\!-\!1)
+\sum_{i=0}^{k-1}\tilde\alpha_{i}\,n^i, \hspace{8mm}
\varphi(n+2)\!=\!\lambda \varphi(n+1) +\rho\, \varphi(n)
+\sum_{i=0}^{k-1}\tilde{\tilde\alpha}_{i}\,n^i. \label{23}
\end{equation}
Validity of these two equations will guarantee fulfillment of the
inhomogeneous Fibonacci relation (\ref{22}) if in addition we
require $\tilde\alpha_{i}+\tilde{\tilde\alpha}_{i}=\alpha_i$.

It can be shown that the ($k+2$)-term\footnote{We count all the terms
in (22), including $\lambda$-term and $\rho$-term.} inhomogeneous FR
is satisfied for all the polynomial oscillators for which $r=k$.
  Inversely, the oscillator with $(r+1)$-th order polynomial
structure function satisfies any $(k+2)$-term inhomogeneous FR
such that \ \ $k\!>\!r$.
On the other hand, the oscillator with $r$-th order polynomial
structure function does not satisfy all the $k$-term generalized
inhomogeneous FRs such that $k\!<\!r$. Accordingly, the $k$-term FR
is not valid for all the polynomial oscillators for which $r\!>\!k$.

Instead of proving the general statements of the latter paragraph we
only give particular examples, the necessary data for which are
placed in the Table.

\begin{center}
\begin{tabular}{|c|l|l|l|}
\hline
{} & {} & {} & {} \\
& Coefficients $\tilde{\alpha}_0, \tilde{\alpha}_1, ...,
\tilde{\alpha}_r$ & Coefficients $\tilde{\tilde\alpha}_0,
\tilde{\tilde\alpha}_1, ..., \tilde{\tilde\alpha}_r$ & Coefficients
$\alpha_0, \alpha_1, ...,
\alpha_r$    \\
{} & from (\ref{23}) & from (\ref{23}) & from (\ref{22}) \\
\hline
{} & {} & {} & {} \\
 $k=1$ & $\tilde{\alpha}_0=2\mu_1$ & $\tilde{\tilde\alpha}_0=2\mu_1$ &
$\alpha_0=4\mu_1$  \\
\hline
{} & {} & {} & {} \\
$k=2$ & $\tilde{\alpha}_0=2\mu_1$ & $\tilde{\tilde\alpha}_0=2\mu_1+6\mu_2$ &
$\alpha_0=4\mu_1+6\mu_2$  \\
{} & $\tilde{\alpha}_1=6\mu_2$ & $\tilde{\tilde\alpha}_1=6\mu_2$ &
$\alpha_1=12\mu_2$  \\
\hline {} & {} & {} & {} \\
$k=3$ & $\tilde{\alpha}_0=2\mu_1+2\mu_3$ &
$\tilde{\tilde\alpha}_0=2\mu_1+6\mu_2+14\mu_3$ &
                                $\alpha_0=4\mu_1+6\mu_2+16\mu_3$  \\
{} & $\tilde{\alpha}_1=6\mu_2$ & $\tilde{\tilde\alpha}_1=6\mu_2+24\mu_3$ &
$\alpha_1=12\mu_2+24\mu_3$  \\
{} & $\tilde{\alpha}_2=12\mu_3$ & $\tilde{\tilde\alpha}_2=12\mu_3$ &
$\alpha_2=24\mu_3$  \\
\hline
{} & {} & {} & {} \\
 $k=4$ & $\tilde{\alpha}_0=2\mu_1+2\mu_3$ &
$\tilde{\tilde\alpha}_0=2\mu_1+6\mu_2$ &
                         $\alpha_0=4\mu_1+6\mu_2$  \\
{} & {} & \ \ \ \ \ $+14\mu_3+30\mu_4$& \ \ \ \ \ {$+16\mu_3+30\mu_4$}\\
 & $\tilde{\alpha}_1=6\mu_2+10\mu_4$ &
$\tilde{\tilde\alpha}_1=6\mu_2+24\mu_3+70\mu_4$ &
                                   $\alpha_1=12\mu_2+24\mu_3+80\mu_4$  \\
{} & $\tilde{\alpha}_2=12\mu_3$ & $\tilde{\tilde\alpha}_2=12\mu_3+60\mu_4$ &
$\alpha_2=24\mu_3+60\mu_4$  \\
{} & $\tilde{\alpha}_3=20\mu_4$ & $\tilde{\tilde\alpha}_2=20\mu_4$ &
$\alpha_2=40\mu_4$  \\
\hline
{} & {} & {} & {} \\
 $k=5$ & $\tilde{\alpha}_0=2\mu_1+2\mu_3+2\mu_5$ &
$\tilde{\tilde\alpha}_0=2\mu_1+6\mu_2$ & $\alpha_0=4\mu_1+6\mu_2$ \\
{} & {} & \ \ \ \ \ {$+14\mu_3+30\mu_4+62\mu_5$} &
\ \ \ \ \ $+16\mu_3+30\mu_4+64\mu_5$\\
{} & $\tilde{\alpha}_1=6\mu_2+10\mu_4$ &
$\tilde{\tilde\alpha}_1=6\mu_2+24\mu_3$ &
                                   $\alpha_1=12\mu_2+24\mu_3+80\mu_4$  \\
{} & {} & \ \ \ \ \ {$+70\mu_4+180\mu_5$} &  \ \ \ \ \ {$+180\mu_5$} \\
{} & $\tilde{\alpha}_2=12\mu_3+30\mu_5$ &
$\tilde{\tilde\alpha}_2=12\mu_3+60\mu_4+210\mu_5$ &
$\alpha_2=24\mu_3+60\mu_4+240\mu_5$  \\
{} & $\tilde{\alpha}_3=20\mu_4$ & $\tilde{\tilde\alpha}_3=20\mu_4+120\mu_5$ &
$\alpha_3=40\mu_4+120\mu_5$  \\
{} & $\tilde{\alpha}_4=30\mu_5$ & $\tilde{\tilde\alpha}_4=30\mu_5$ &
$\alpha_4=60\mu_5$  \\
\hline
\end{tabular}
\end{center}

The quadratically deformed oscillator, with $r=1$ or
$\varphi(n)=n+\mu_1n^2$,
 does not satisfy the standard FR
(\ref{1}), but it obeys the simplest inhomogeneous FR
\begin{equation}
E_{n+1}=\lambda E_n+\rho E_{n-1}+\alpha_0 \ , \hspace{10mm}
\lambda=2, \hspace{4mm} \rho=-1 , \hspace{5mm} \alpha_0=4\mu_1, \label{24}
\end{equation}
see the $k=1$ row in the Table.
 Let us note that, whatever is $k$ (i.e., for any power in $n$ of the
polynomial structure function), the coefficients $\lambda$, $\rho$
will be always $\lambda=2$, $\rho=-1$.
 The set $\alpha_0, \alpha_1,...,$ however, differs for different $k$,
as seen in the five rows of the Table.

Remark that, contrary to the case of $k$-bonacci relation where all
the coefficients $\lambda_i^{(k)}$ in (\ref{14}) are really
constant (independent of $n$ {\em and} $\{\mu\}$), here $\alpha_i$
are functions of $\mu_j$.

\subsection{Polynomially deformed oscillators as quasi-Fibonacci ones}
\hspace{3mm}
  In subsection 2.2 we assumed the coefficients
$\lambda_i$, $i=0,...,k-1$, in the $k$-generalized Fibonacci (or
$k$-bonacci) relation (\ref{13}) to be real constants.
 In this subsection we modify the initial two-term linear,
{\bf standard} FR (\ref{1}) by admitting an explicit dependence
on the number $n$ of both $\lambda$ and $\rho$ entering the
relation.
 That is, now we deal with the
so-called {\it quasi-Fibonacci relation}\footnote{Below, for
convenience, we denote $\lambda(n)$ and $\rho(n)$ also as $\lambda_n$ and
$\rho_n$.}:
\begin{equation}
E_{n+1}=\lambda(n)E_n+\rho(n)E_{n-1}\ . \label{25}
\end{equation}

Let us note that for the $\mu$-oscillator from \cite{Jann}, which is
non-Fibonacci, its quasi-Fibonacci properties have been described in
detail in Ref. \cite{GKR} where three different ways of deriving
$\lambda_n$ and $\rho_n$ have been explored. Here, for the
polynomially deformed or $\{\mu\}$-oscillators, only two of the
three are considered.

Following the first way
 we deal with the system of equations related with (\ref{25}), namely
\begin{equation}
\begin{cases} \varphi(n+1)=\lambda_n \varphi(n)+\rho_n \varphi(n-1)\ ; \cr
\varphi(n+2)=\lambda_n \varphi(n+1)+\rho_n \varphi(n)\ . \label{26}
\end{cases}
\end{equation}
Simultaneous validity of them both guarantee fulfillment of (\ref{25}).
Solving of (\ref{26}) yields
\begin{equation}
\lambda_n=\frac{\varphi(n+1)-\rho_n\,\varphi(n-1)}{\varphi(n)}\ ,
\hspace{5mm} \rho_n=\frac{\varphi(n+2)\varphi(n)-\varphi^2(n+1)}
{\varphi^2(n)-\varphi(n+1)\varphi(n-1)}\ .  \label{27}
\end{equation}
With account of the explicit form (\ref{5})
of the structure function we have
\[
\rho_n=\frac{\sum_{i=0}^k\mu_in^{i+1}\sum_{j=0}^k\mu_j(n+2)^{j+1}-
\sum_{i=0}^k\mu_i(n+1)^{i+1}\sum_{j=0}^k\mu_j(n+1)^{j+1}}
{\sum_{i=0}^k\mu_in^{i+1}\sum_{j=0}^k\mu_jn^{j+1}-
\sum_{i=0}^k\mu_i(n-1)^{i+1}\sum_{j=0}^k\mu_j(n+1)^{j+1}}.
\]
The obtained expression for $\rho_n$ by plugging it in eq. (\ref{27})
yields also $\lambda_n$.

To proceed in the second way, see \cite{GKR}, we put $
\rho_n=\lambda_{n-1}$ in (\ref{12}), that gives
\[
E_{n+1}=\lambda_{n}E_{n}+\lambda_{n-1}E_{n-1}
\]
or
\begin{equation}
\lambda_{n+1}+\frac{E_n}{E_{n+1}}~\lambda_n=\frac{E_{n+2}}{E_{n+1}}\
,
 \hspace{5mm} n\geq 0 \ . \label{28}
\end{equation}
 With the initial condition $\lambda_0=c$, we find by induction the formula
\[
\lambda_n\equiv\lambda(n)=\frac{\sum_{j=2}^{n+1}(-1)^{n-j+1}E_j+(-1)^ncE_0}{E_n}
\]
which in terms of the structure function looks as
\[
\lambda_n=\frac{\sum_{j=2}^{n+1}(-1)^{n-j+1}\varphi(j)+(-1)^nc\varphi(0)}{\varphi(n)}\
.
\]
With account of (\ref{5}), the expressions for $\lambda_n$
 and $\rho_n=\lambda_{n-1}$ which provide validity of the
 quasi-Fibonacci relation (\ref{25})
 take the final explicit form
\begin{equation}
\lambda_n=\frac{\sum_{j=2}^{n+1}(-1)^{n-j+1}\sum_{i=0}^s\mu_ij^{i+1}}
{\sum_{i=0}^s\mu_in^{i+1}}\ , \hspace{5mm}
\rho_n=\frac{\sum_{j=2}^{n}(-1)^{n-j}\sum_{i=0}^s\mu_ik^{i+1}}
{\sum_{i=0}^s\mu_i(n-1)^{i+1}}\ . \label{29}
\end{equation}
This completes our short quasi-Fibonacci treatment of the polynomially
deformed $\{\mu\}$-oscillators, \ ${\mu}\equiv(\mu_1,
\mu_2,...,\mu_n)$.

\section{Deformed oscillators, polynomial in  $q$- or
$p,\!q$-brackets}\hspace{5mm}

Here we examine some other, than the ${\{\mu\}}$-deformed, classes of
oscillators (with added more canonical deformation parameters) obeying Tribonacci
and higher order relations.

\subsection{
 A class of $(q;\mu)$\,- and $(q;\{\mu\})$\,-deformed
oscillators}
  Consider the $(q;\mu)$-deformed oscillator defined by
the structure function
\begin{equation} \varphi_{n}(q;\mu)\equiv\varphi(n;
q,\mu)=[n]+\mu [n]^2=[n]\bigl(1+\mu[n]\bigr)\ , \label{30}
\end{equation}
\begin{equation}
[n]\equiv[n]_q=\frac{1-q^n}{1-q}, \hspace{5mm} q>0 \ .\label{31}
\end{equation}
It can be proven that the structure function (\ref{30}) and
hence the energy values $E_{n}$ of such oscillators obey the 3-term
extended Fibonacci (= Tribonacci) relation
\begin{equation}
\varphi_{n+1}(q,\mu)= \lambda(q)\, \varphi_n(q,\mu) +
\rho(q)\,\varphi_{n-1}(q,\mu) + \sigma(q)\,\varphi_{n-2}(q,\mu)\ ,
\label{32}
\end{equation}
\begin{equation}
\hspace{9mm}  E_{n+1}= \lambda(q) E_{n} + \rho(q) E_{n-1} +
\sigma(q) E_{n-2}\ , \label{33}
\end{equation}
where $\lambda(q)$, $\rho(q)$, $\sigma(q)$ depend on the parameter
$q$ as
\[
\lambda(q) =[3] \ , \hspace{12mm} \rho(q) =-q[3] \ , \hspace{12mm}
\sigma(q) = q^3 \
\]
(compare with (\ref{12})
and its coefficients $\lambda_0=3,\ \lambda_1=-3,\ \lambda_2=1$).
 The result in (\ref{30}), (\ref{32})-(\ref{33}), with
the above $\lambda(q)$, $\rho(q)$, $\sigma(q)$, generalizes to the
following statement.

\vspace{2mm}
 {\bf Proposition 2.} For the $(q;\{\mu\})$-deformed oscillators,
their structure function
\begin{equation}
\varphi_n(q;\{\mu\})=[n]_q+\sum^r_{j=1}\mu_j\, ([n]_q)\!^{j+1} \  ,
\hspace{8mm} \{\mu\}\equiv(\mu_1,\mu_2,...,\mu_r),      \label{34}
\end{equation}
and thus the energies $E_{n}$ obey the $k$-term extended Fibonacci
(or "$k$-bonacci") relation
\begin{equation}
\varphi_{n+1}(q;\{\mu\})= \lambda_0\,\varphi_n(q;\{\mu\}) +
\lambda_1\,\varphi_{n-1}(q;\{\mu\}) + ... +
\lambda_{k-1}\,\varphi_{n-k+1}(q;\{\mu\})\ , \label{35}
\end{equation}
\begin{equation}
\hspace{9mm}  E_{n+1}= \lambda_0\,E_n(q;\{\mu\}) +
\lambda_1\,E_{n-1}(q;\{\mu\}) + ... +
\lambda_{k-1}\,E_{n-k+1}(q;\{\mu\})\ , \label{36}
\end{equation}
if $r=k-2$ and the coefficients $\lambda_i=\lambda_i(q)$,
$i=0,1,...,k-1$, are taken in the form
\begin{equation}
\lambda_i(q)=\lambda^{(k)}_i(q)=(-1)^i\,
q^{i(i+1)/2}\frac{[k]!}{[i+1]![k-1-i]!}=
                (-1)^i\,q^{i(i+1)/2}\left(
                             \begin{array}{c}
                                  \hspace{-2mm}k\hspace{-2mm}\\
                                 \hspace{-2mm}i\!+\!1\hspace{-2mm}\\
                            \end{array}
                                  \right)_q\ .    \label{37}
\end{equation}
The proof proceeds in analogy with that of Proposition 1.

Note: in the limit $q\rightarrow 1$, formulas (\ref{34}) and
(\ref{37}) reduce respectively to (\ref{5}) and (\ref{14}), that in
this limit gives complete recovery of the Proposition 1. As an
 interesting fact let us stress the independence of $\lambda_i^{(k)}(q)$
 on $\{\mu\}$ in (\ref{37}), and in this limit.

\subsection{A class of ($p,q;\mu$)-deformed nonlinear
oscillators}\hspace{5mm}

 Let us remind that the $q$-deformed oscillator linear
in the $q$-bracket (\ref{31}) possesses, from one hand, the standard
2-term Fibonacci property (1) while, from the other hand, is a
particular $p=1$ case of the $p,\!q$-deformed oscillator whose
$p,\!q$-bracket is
\begin{equation}
[n]_{p\!,q}\equiv\frac{p^n-q^n}{p-q} \ . \label{38}
\end{equation}
 We could expect that the 3-term (Tribonacci) relation holds
for the oscillators involving besides $\mu$ two more deformation
parameters $p$, $q$, so that the structure function is
\begin{equation}
\varphi_{n}(p,q;\mu) =[n]_{p\!,q}+\mu
[n]_{p\!,q}^2=[n]_{p\!,q}\bigl(1+\mu[n]_{p\!,q} \bigr)\ . \label{39}
\end{equation}
But, it turns out this fails. To prove this fact suppose the
opposite, that deformed oscillator with the structure
function (\ref{39}) obeys the Tribonacci relation
\begin{equation}
\varphi_{n+1}(p,q;\mu)= \lambda({p,q})~ \varphi_n(p,q;\mu) +
\rho({p,q})~ \varphi_{n-1}(p,q;\mu) + \sigma({p,q})~
\varphi_{n-2}(p,q;\mu)\ . \label{40}
\end{equation}
Insert (\ref{39}) into the relation (\ref{40}) and
by equating the corresponding coefficients deduce the following set
of equations:

\vspace{3mm} $p^n$:\ \ \ \ \hspace{5mm}
$p^2-pq=\lambda(p-q)+\rho(1-p^{-1}q)+\sigma(p^{-1}-p^{-2}q)$\ ,

\vspace{3mm} $q^n$:\ \ \ \ \hspace{5.5mm}
$q^2-pq=\lambda(q-p)+\rho(1-q^{-1}p)+\sigma(q^{-1}-q^{-2}p)$\ ,

\vspace{3mm} $(pq)^n$:\ \ \ \ \ $pq=\lambda+\rho p^{-1}q^{-1}+\sigma
p^{-2}q^{-2}$\ ,

\vspace{3mm} $p^{2n}$:\ \ \ \ \hspace{4mm} $p^2=\lambda+\rho
p^{-2}+\sigma p^{-4}$\ ,

\vspace{3mm}  $q^{2n}$:\ \ \ \ \hspace{4mm} $q^2=\lambda+\rho
q^{-2}+\sigma q^{-4}$\ . \ \ \ \

\vspace{3mm}\noindent
 This system of equations is easily shown to be inconsistent (having
no solutions).
 Therefore, the structure function (\ref{39}) does not satisfy
the Tribonacci relation (\ref{40}).

Analogously to this negative result, it can be proven that the
deformed oscillator under question does not satisfy as well the 4-term (or
Tetranacci) relation.

In the next subsection we will show how to properly treat the
deformed oscillators defined by the structure function (\ref{39})
and alike, from the viewpoint of yet higher extension of the
Fibonacci (Tribonacci, Tetranacci) relations.

\subsection{Deformed $(p,q;\mu)$-oscillators as Pentanacci oscillators}
\hspace{3mm}

One can prove that the following statement is true.

{\bf Proposition 3.} The family of $(p,\!q;\mu)$-oscillators
 with quadratic in $[n]_{p,q}$ structure function (\ref{39})
obeys the Pentanacci (5-term extended
Fibonacci) relation
\begin{equation}
\varphi_{n+1}=\lambda(p,q)\varphi_{n}+\rho(p,q)\varphi_{n-1}+
\sigma(p,q)\varphi_{n-2}+\gamma(p,q)\varphi_{n-3}+\delta(p,q)\varphi_{n-4}\
\label{41}
\end{equation}
if the coefficients $\lambda(p,q), \rho(p,q), \sigma(p,q),
\gamma(p,q), \delta(p,q)$ are\footnote{Note their
$\mu$-independence.}:
\[\lambda(p,q)=p^2+q^2+p+q+pq= [2]_{p,q}+[3]_{p,q},\]
\[\hspace{-7mm}\rho(p,q)=-p^3q-p^3-2p^2q-p^2q^2-pq^3-2pq^2-pq-q^3=\]
\[\hspace{5.5mm}=-\Bigl([4]_{p,q}+pq\bigl(1+[2]_{p,q}+[3]_{p,q}\bigr)\Bigr)=
-\Bigr([3]_{p,q}\cdot[2]_{p,q}+pq(1+[3]_{p,q})\Bigl),\]
\[\sigma(p,q)=2p^3q^2+2p^2q^3+p^4q+pq^4+p^3q+pq^3+p^2q^2+p^3q^3=
pq\Bigr([3]_{p,q}([2]_{p,q}+1)+p^2q^2\Bigl),
\]
\[\hspace{-7mm}\gamma(p,q)=-\bigl(p^2q+p^2+pq^2+pq+q^2\bigr)p^2q^2=
-p^2q^2\bigr([3]_{p,q}+pq[2]_{p,q}\bigl),\]
\begin{equation}
\hspace{-7mm}\delta(p,q)=p^4q^4. \label{42}
\end{equation}
The {\it proof} proceeds by direct verification.

{\bf Remark 3.} One can show that these same oscillators satisfy
also respective $k$-term extended Fibonacci relation for an integer
$k\geq5$. Let us illustrate this for $k=6$.
 We take one more copy of the relation (\ref{41}) in which the shift
$n\to n-1$ is done, and subtract this copy, multiplied by some
$\kappa$, from the initial relation (\ref{41}).
 Then the 6-term extended relation
\[
\varphi_{n+1}=(\lambda-\kappa)\,\varphi_{n}+(\rho+\lambda\kappa)\,\varphi_{n-1}+
(\sigma+\rho\kappa)\,\varphi_{n-2}+(\gamma+\sigma\kappa)\,\varphi_{n-3}\,+
\]
\[
\hspace{10mm} +\,(\delta+\gamma\kappa)\,\varphi_{n-4}+
\kappa\,\varphi_{n-5}
\]
results and is valid for any real number $\kappa$, with the
coefficients $\lambda,  \rho, \sigma, \gamma, \delta$ taken from
(\ref{42}).
  Note, this same procedure can be applied any desired
number of times, with the appropriate shifts $n\rightarrow n-j$
(clearly, $j<n$).

{\bf Remark 4.} It is of interest to check the $p=1$ limit of
(\ref{39}) and (\ref{41})-(\ref{42}). Contrary to naive expectation
that we should obtain the $r=5$ case of Proposition 2, with the
coefficients $\lambda_i^{(5)}$, given in (\ref{37}), we find a kind
of surprise: the coefficients that follow from (\ref{42}) are other
than $\lambda_0^{(5)}, \lambda_1^{(5)}, \lambda_2^{(5)},
\lambda_3^{(5)}$ and $\lambda_4^{(5)}$. This "controversy"\ is
rooted in the fact that the $(q;\mu)$-oscillator with
$\varphi(n)=[n]_q+\mu[n]_q^2$ respects already the Tribonacci
(3-term extended Fibonacci) relation while usual FR fails.  The
situation with $(p,q;\mu)$-oscillator is more involved: for it, both
the usual 2-term FR and the 3-, 4-term extensions of Fibonacci
relations are not valid; only the 5-term or Pentanacci relation does
hold.

\subsection{Nine-bonacci deformed $(p,q;\mu_1,\mu_2)$-oscillators}
 \hspace{3mm}
 Consider the cubic in the $p,\!q$-bracket $[n]_{p,q}$
structure function
\begin{equation}
\varphi_n=[n]_{p,q}+\mu_1[n]_{p,q}^2+\mu_2[n]_{p,q}^3\ , \hspace{8mm}
[n]_{p,q}=\frac{p^n-q^n}{p-q}\ . \label{43}
\end{equation}
 It can be proven that such oscillator does not satisfy the standard
FR, nor it satisfies any $k$-term extended, $k\leq
8$, $k$-bonacci relation.
 However, it does satisfy the 9-term extended version of FR
({\em "Nine-bonacci relation"}), as reflected in the next statement.

{\bf Proposition 4}. The $(p,q;\mu_1,\mu_2)$-oscillator given by the
structure function $\varphi(n)$ in (\ref{43}) satisfies the $9$-term
extension of FR or Nine-bonacci relation of the form
\begin{equation}
\varphi_{n+1}=\sum_{j=0}^8A_j\,\varphi_{n-j} \label{44}
\end{equation}
if the coefficients $A_j\equiv A_j(p,q)$ are given as ({\em note
their} $\mu_1,\mu_2$\,-{\em independence})
\[
A_0(p,q)=[4]_{p,q}+[3]_{p,q}+[2]_{p,q},
\]
\[
A_1(p,q)=-([6]_{p,q}+(1+pq)[5]_{p,q}+(1+pq)[4]_{p,q}+2pq[3]_{p,q}+\]
\[\hspace{15mm}+pq(1+pq)[2]_{p,q}+pq(1+p^2q^2)),
\]
\[
A_2(p,q)=(1+pq)[7]_{p,q}+2pq[6]_{p,q}+pq(2+pq)[5]_{p,q}+pq(2+4pq+p^2q^2)[4]_{p,q}+\]
\[\hspace{15mm}+pq(1+2pq+2p^2q^2)[3]_{p,q}+pq(pq+2p^2q^2)[2]_{p,q}+2p^3q^3,
\]
\[
A_3(p,q)=-pq\Bigl([8]_{p,q}+(1+pq)[7]_{p,q}+
(1+2pq+p^2q^2)[6]_{p,q}+pq(3+2pq)[5]_{p,q}+\]
\[\hspace{22mm}+pq(1+4pq+p^2q^2)[4]_{p,q}+pq(1+2pq+3p^2q^2)[3]_{p,q}+
p^2q^2(2+2pq+p^2q^2)[2]_{p,q}+\]
\[\hspace{15mm}+p^3q^3(2+p^2q^2)\Bigr),
\]
\begin{equation}
A_4(p,q)=p^2q^2\Bigl([8]_{p,q}+(1+pq)[7]_{p,q}+
(1+2pq+p^2q^2)[6]_{p,q}\Bigr)+ \label{45}
\end{equation}
\[
\hspace{25mm}+p^3q^3\Bigl((2+3pq)[5]_{p,q}+(1+4pq+p^2q^2)[4]_{p,q}\Bigr)
+p^4q^4\Bigl((3+2pq+p^2q^2)[3]_{p,q}+
\]
\[\hspace{15mm}+(1+2pq+2p^2q^2)[2]_{p,q}+(1+2p^2q^2)\Bigr),
\]
\[ \hspace{10mm}
A_5(p,q)=-p^3q^3\Bigl((1+pq)[7]_{p,q}+2pq[6]_{p,q}+pq(1+2pq)[5]_{p,q}+
pq(1+2pq+2p^2q^2)[4]_{p,q}+\]
\[\hspace{15mm}+p^2q^2(2+2pq+p^2q^2)[3]_{p,q}+p^3q^3(2+pq)[2]_{p,q}+2p^4q^4\Bigr),
\]
\[
A_6(p,q)=p^7q^7\Bigl([6]_{p,q}+
2[3]_{p,q}+(1+pq)[2]_{p,q}+(2+p^2q^2)\Bigr)\]
\[\hspace{15mm}+p^5q^5(1+pq)[5]_{p,q}+p^6q^6(1+pq)[4]_{p,q},
\]
\[
A_7(p,q)=-p^7q^7\Bigl([4]_{p,q}+pq[3]_{p,q}+p^2q^2[2]_{p,q}\Bigr),
\]
\[A_8(p,q)=p^{10}q^{10}.
\]
 The proof is achieved by direct verification.

{\bf Remark 5.} If $p\rightarrow 1$, (\ref{43}) reduces to the
$r=2$ case of (\ref{34}), and the coefficients (\ref{45}) turn
into
\[
A_0(q)=[4]_q+[3]_q+[2]_q,
\]
\[
A_1(q)=-2q^2([4]_q+[3]_q+2q)-q^2[2]_q,
\]
\[
A_2(q)=[8]_q+5q[6]_q+2q[5]_q+q[4]_q+6q^2[3]_q+q^3[2]_q+6q^3,
\]
\[
A_3(q)=-q(3[8]_q+6q[6]_q+2q[5]_q+5q^2[4]_q+6q^3[3]_q+8q^3[2]_q+2q^4),
\]
\begin{equation}
A_4(q)=3q^2[8]_q+6q^3[6]_q+2q^4[5]_q+11q^4[4]_q+4q^5[2]_q+4q^6,
\label{46}
\end{equation}
\[
A_5(q)=-q^3([8]_q+6q[6]_q+q[5]_q+4q^2[4]_q+3q^3[3]_q+3q^3[2]_q+4q^4,
\]
\[
A_6(q)=2q^5[6]_q+q^6[5]_q+q^6[4]_q+2q^7[3]_q+3q^8[2]_q+3q^9,
\]
\[
A_7(q)=-q^7([4]_q+q[3]_q+q^2[2]_q),
\]
\[
A_8(q)=q^{10}.
\]
Here again we find a surprise: the $p\rightarrow 1$ limits of the
coefficients $A_j$ do not merge with those stemming   
from the general formula (\ref{37}). This is rooted in the fact that
the structure function (\ref{43}) goes over into
$\varphi(n)=[n]_q+\mu_1[n]^2_q+\mu_2[n]_q^3$, for which already the
Pentanacci relation is valid as follows from Proposition 2.
 To lift the controversy, let us show that it is possible to derive
the 9-term extended FR (\ref{44}), just with the coefficients
(\ref{46}), starting from the Pentanacci relation given by
(\ref{41})-(\ref{42}) at $k=4$. Indeed, consider besides the
Pentanacci relation (\ref{41}) at $k=4$, with fixed $n$, the four
additional copies of it written accordingly other shifts $n\to n-1,
n\to n-2, n\to n-3, n\to n-4$.
 Then multiply these four relations respectively by $t, x, y, z$ of the form
\[t=-([2]+[3]),\]
\[
x=-(2q^2+1)([4]+[3])-[2](q^2-1)-4q^3,
\]
\[
y=-[8]-5q[6]\!-\!2q[5]+[4](2q^2\!-\!q+2)+[3](-4q^2+2)+[2](-q^3+q^2\!-\!1)\!-\!2q^3,
\]
\[
z=3q[8]+[6](6q^2+5q)+2q[5](q+1)+[4]q(5q^2+1)+[3](6q^4+6q^2+4)+
\]
\[
\hspace{3.5mm}+\,[2](8q^4+q^3+1)+2q^5+6q^3,
\]
and take their sum, term by term, with the first copy. That will
lead to the above Nine-bonacci relation (\ref{44}), with exactly
the coefficients given in (\ref{46}).

 We conclude this section with the {\em remark concerning
general situation} of deformed oscillators with a polynomial of any order in
$[n]_{p,q}$ structure function, cf. (\ref{43}).
 It can be argued that the oscillator for which $\varphi(n)$ is
$(k+1)$-th order polynomial in $[n]_{p,q}$ obeys the $m$-bonacci
relation where $m=\frac12(k+2)(k+3)-1$. It is however hardly
possible to find the coefficients $\lambda_0, \lambda_1,
\lambda_2,...,\lambda_{m-1}$ in explicit form {\it for arbitrary}
$m$.

\section{Conclusions}\hspace{3mm}
In this paper we have shown, first, that deformed oscillators whose
structure function is a polynomial in $n$, or $[n]_q$, or
$[n]_{p,q}$, do not belong to the Fibonacci class.
 On the other hand, the $\{\mu\}$-oscillators
(respectively $(q;\{\mu\})$-oscillators) for which $\varphi(n)$ is
polynomial in $n$ (respectively polynomial in $[n]_q$) share the
characteristic property: they satisfy the $k$-bonacci relation, with
the coefficients (\ref{14}) or (\ref{37}), if the order of
polynomial is $r=k-2$.
 In particular, the Tribonacci relation occurs if the structure function
is quadratic.
 It is worth to mention a remarkable fact that in the both these cases
 the coefficients (\ref{14}) and (\ref{37}) of the $k$-bonacci relation
 {\em do not depend on the set of numbers} ${\mu_i}$ involved in the polynomial,
 and depend only on its label (=subscript) and on the number $k$, linked as
 $r=k-2$ to the order of the polynomial.
 Independence of $\mu$ or $\mu_1$, $\mu_2$ is also seen in (\ref{42}) and (\ref{45})
 for the Pentanacci and Nine-bonacci relations.
 Instead, the "initial values"\ $E_0$, $E_1$,..., $E_{k-1}$ for the $k$-bonacci
 relation {\it inevitably depend} on $\mu_1$, $\mu_2$,..., $\mu_{k-2}$, as it is given by the
 formula (\ref{3}) for $E_n$ joint with (\ref{5}); see also (\ref{35}), (\ref{36}).

  Moreover, for deformed $\mu$-oscillators whose $\varphi(n)$ is given in (\ref{5})
 we have studied the alternative possibility that the deformed oscillators of this family
 may also be considered as those obeying the inhomogeneous FR, see (\ref{22}).
 However, in this case, unlike the already mentioned property of $\lambda_i^{(k)}$ and
 $\lambda_i^{(k)}(q)$ in (\ref{14}) and (\ref{37}), the coefficients $\tilde{\lambda_i}$
 and $\tilde{\tilde{\lambda_i}}$ and $\lambda_i$ from (\ref{22}), (\ref{23}) {\it do depend}
 on the numbers ${\mu_i}\equiv(\mu_1,...,\mu_2)$ which determine the polynomial
structure function (\ref{5}).
 This fact is manifest in the Table of subsection 2.3.
 Concerning the related class of deformed oscillators, polynomial in $[n]_q$, one can also
 consider them from the viewpoint of inhomogeneous FR, but with important change: the
sum appearing in the RHS of the analogue of (\ref{23}) should now be
taken over the powers of $q^n$ (instead of the powers of $n$).
 We have also shown that deformed oscillators of these two classes can
be treated as quasi-Fibonacci ones, i.e. those obeying the relation
with 2-term RHS where $\lambda=\lambda(n)$ and $\rho=\rho(n)$.

As even more unusual looks the situation with the class of deformed
oscillators whose $\varphi(n)$ is polynomial in the $p,\!q$-bracket
$[n]_{p,q}$. Indeed, the structure function quadratic in $[n]_{p,q}$
obeys the Pentanacci relation first while the FR, Tribonacci and
Tetranacci relations all fail. Then, $\varphi(n)$ cubic in
$[n]_{p,q}$ defines the family of oscillators which are 9-bonacci
oscillators, and so on, according to the rule: $m$-bonacci relation
corresponds to $(k+1)$-th order polynomial, in $[n]_{p,q}$,
deformation where $m=\frac12(k+2)(k+3)-1$. Finally, let us note that
the both, $(q;\{\mu\})$- and $(p,q;\{\mu\})$-deformed oscillators
can be viewed as quasi-Fibonacci oscillators, in complete analogy
with our above treatment (in sec. 3.2) of $\{\mu\}$-oscillators, and
in analogy with the content of our work \cite{GKR}.

Our final remarks concern the issue of physics aspects of the
polynomially deformed non-Fibonacci oscillators, in the one-mode
(non-covariant) case studied in this paper.
 We believe these nonlinear oscillators have good potential to find effective
 applications in a number of quantum-physics branches, from quantum optics \cite{Man',Alvarez,Sunil}
  to deformed field theory, say, in the spirit of \cite{Man'2,Curado}.
  About special manifestations of just the non-Fibonacci nature of
  deformed oscillators,     
  at present we can only mention our recent work \cite{GR_mu-B} on the application of $\mu$-oscillators
  (being not Fibonacci but quasi-Fibonacci, see \cite{GKR}) for
  constructing respective $\mu$-Bose gas model in analogy with the $p,q$-Bose gas
  model treated in \cite{AdGa}. Unlike the latter, for $\mu$-Bose gas the evaluations of (intercepts of)
  2-, 3-particle correlations are significantly more involved and do not yield closed
  expressions: only approximate formulas can be obtained.

\subsection* {Acknowledgements}

The authors are thankful to I.M. Burban and I.I. Kachurik for
valuable discussions. This research was partially supported by the
Grant 29.1/028 of the State Foundation of Fundamental Research of
Ukraine, and by the Special Program of the Division of Physics and
Astronomy of the NAS of Ukraine.


\end{document}